\documentclass[aps, pra,nobalancelastpage,twocolumn,superscriptaddress,nolongbibliography]{revtex4-1}
\pdfoutput=1

\usepackage{color,amsthm,amsmath,amsxtra,amsfonts,dsfont,graphicx,bm,amssymb}

\usepackage[colorlinks=true,linkcolor=blue, citecolor=blue, urlcolor=blue, bookmarks]{hyperref}
\usepackage{centernot}
\usepackage{subfigure}

\usepackage[dvipsnames]{xcolor}

\usepackage{tensor}

\include{new_commands}

\def\be{\begin{equation}}
	\def\ee{\end{equation}}
\def\bea{\begin{eqnarray}}
	\def\eea{\end{eqnarray}}

\usepackage[normalem]{ulem}

\newcommand{\ket}[1]{\vert#1\rangle}
\newcommand{\bra}[1]{\langle#1\vert}

\begin{document}

\title{Symmetry-resolved Page curves}

\author{Sara Murciano}
\affiliation{SISSA and INFN Sezione di Trieste, via Bonomea 265, 34136 Trieste, Italy}

\author{Pasquale Calabrese}
\affiliation{SISSA and INFN Sezione di Trieste, via Bonomea 265, 34136 Trieste, Italy}
\affiliation{The Abdus Salam International Center for Theoretical Physics, Strada Costiera 11, 34151 Trieste, Italy}

\author{Lorenzo \surname{Piroli}}
\affiliation{Philippe Meyer Institute, Physics Department, \'Ecole Normale Sup\'erieure (ENS), Universit\'e PSL, 24 rue Lhomond, F-75231 Paris, France}

\begin{abstract}

Given a statistical ensemble of quantum states, the corresponding \emph{Page curve} quantifies the average entanglement entropy associated with each possible spatial bipartition of the system. In this work, we study a natural extension in the presence of a conservation law and introduce the \emph{symmetry-resolved} Page curves, characterizing average bipartite symmetry-resolved entanglement entropies. We derive explicit analytic formulae for two important statistical ensembles with a $U(1)$-symmetry: Haar-random pure states and random fermionic Gaussian states. In the former case, the symmetry-resolved Page curves can be obtained in an elementary way from the knowledge of the standard one. This is not true for random fermionic Gaussian states. In this case, we derive an analytic result in the thermodynamic limit based on a combination of techniques from random-matrix and large-deviation theories. We test our predictions against numerical calculations and discuss the sub-leading finite-size corrections. 

\end{abstract}

\maketitle

\section{Introduction}

Consider an isolated quantum system $S$ in a random pure state $\ket{\psi}$. Taking a bipartition $S=A\cup B$, one may ask what is the corresponding \emph{entanglement entropy}~\cite{nielsen2002quantum}. The answer is encoded in an elegant formula conjectured by Page~\cite{page1993average}, now a classical result in quantum mechanics.

Page's contribution is one of the first steps towards a systematic characterization of entanglement in random quantum states, with important applications in different fields, ranging from the black-hole information paradox~\cite{page1993information,hayden2006aspects,hayden2007black,almheiri2020,stanford2022} to foundations in statistical mechanics~\cite{popescu2006entanglement,gogolin2016equilibration}. 

From the point of view of many-body physics,  Page's formula provides qualitative insight into generic systems, as random states are expected to capture the behavior of either eigenstates of typical Hamiltonians~\cite{dalessio2016quantum,vidmar2017entanglement_II,nakagawa2018universality,fujita2018page,lu2019renyi,bianchi2021volume}, or of states generated by a sufficiently chaotic dynamics~\cite{brandao2016local,onorati2017mixing}. In addition, it shows the power of random ensembles to characterize entanglement-related quantities. This is particularly interesting, given the established difficulty to study them in many-body systems~\cite{amico2008entanglement,calabrese2009entanglement,eisert2010colloquium}.  

In his work, Page considered pure states distributed according to the Haar measure and focused on the averaged von Neumann entanglement entropy $S_1$, describing what is now known as the \emph{Page curve}. Although it is non-trivial, its form becomes elementary in the limit of large Hilbert-space dimensions: Given the Hilbert space $\mathcal{H}=\mathcal{H}_A\otimes \mathcal{H}_B$ associated with the system bipartition $S=A\cup B$, Page's formula gives 
\be
S_1= {\rm min}(\log d_A, \log d_B)+O(1),
\ee
where $d_A$, $d_B$ are the dimensions of $\mathcal{H}_A$, $\mathcal{H}_B$. That is, up to subleading corrections, the average entanglement entropy of a subsystem is the maximal one, meaning that typical states display almost maximal bipartite entanglement.

Page's formula, which built upon earlier results~\cite{lloyd1988complexity,lubkin1993average}, was later proven in Refs.~\cite{foong1994proof,sanchez1995simple,sen1996average, zyczkowski2001induced}, and inspired generalizations in a series of works characterizing higher entanglement moments~\cite{cappellini2006distribution,giraud2007distribution, vivo2016random, wei2017proof,bianchi2019typical}, large deviations~\cite{facchi2008phase, nadal2010phase,nadal2011statistical,depasquale2010phase,chen2010smallest,facchi2013entropy, facchi2019phase,bianchi2019typical, nakata2020generic,morampudi2020universal}, entanglement spectrum~\cite{vznidarivc2006entanglement,majumdar2008exact,deelan2013typical},  mixed-state purity ~\cite{depasquale2011statistical} and negativity~\cite{hejazi2021symmetry,shapourian2021entanglement,jaydeep2022page,kudler2022negativity}. Recently, some of these results were also extended to ensembles of \emph{random Gaussian states}, capturing the typical behavior of \emph{non-interacting} systems. In particular, motivated by studies of entanglement in random spin chains~\cite{vidmar2017entanglement,vidmar2018volume,hackl2019average,lydzba2020eigenstate,lydzba2021entanglement}, the Page curve for fermionic Gaussian states was computed for finite systems in Ref.~\cite{bianchi2021page}, while further results were obtained in the thermodynamic limit~\cite{liu2018quantum,pengfei2020subsystem,bernard2021entanglement}, see Ref.~\cite{bianchi2021volume} for a review. 

In this context, an important problem is the characterization of entanglement in random ensembles displaying a conserved quantity, a setting which resembles more closely typical many-body systems. While a number of previous studies have focused on the standard entanglement entropy~\cite{bianchi2019typical,hejazi2021symmetry, bianchi2021volume, bernard2021entanglement,nakata2020generic}, a very natural question pertains to its \emph{symmetry-resolved} (SR) version~\cite{laflorencie2014spin,moshe2018symmetry,xavier2018equipartition}. This quantity, which has been recently the subject of an intense research activity~\cite{bonsignori2019symmetry,belin2013holographic,caputa2016charged,dowker2016,dowker2017,cornfeld2018imbalance,
barghathi2018renyi,barghathi2019operationally,feldman2019dynamics,cornfeld2019entanglement,
fraenkel2020symmetry,calabrese2020full,monkman2020operational,azses2020symmetry,azses2020identification,
barghathi2020theory,tan2020particle,murciano2020symmetry,turkeshi2020entanglement,kiefer2020bounds,
murciano2020symmetry_2D,kiefer2020evidence,capizzi2020symmetry,murciano2020entanglement,horvath2020symmetry,
bonsignori2020boundary,estienne2021finite,murciano2021symmetry,
chen2021symmetry,zhao2021symmetry,weisenberger2021symmetry,
capizzi2021symmetry,hung2021entanglement,calabrese2021symmetry,
horvath2021u,azses2021observation,kiefer2021slow,shachar2021entanglement,parez2021exact,parez2021quasiparticle,ma2022symmetric,
oblak2022equipartition,zhao2022symmetry,ares2022symmetry,jones2022symmetry,
horvath2022branch,chen2022charged,ghasemi2022universal,scopa2022exact,parez2022dynamics,chen2022negativityboson,multicharged,goldstein2022}, is experimentally accessible~\cite{lukin2019probing,vitale2022symmetry} and its behavior encodes interesting information about the interplay between symmetry and entanglement.

In this work, we investigate the SR entanglement entropy in the two statistical ensembles discussed above: Haar-random and fermionic Gaussian random states, with an additional $U(1)$ symmetry. We derive explicit analytic formulae for the corresponding \emph{symmetry-resolved Page curves}. First, following Ref.~\cite{bianchi2019typical}, we show that the latter can be derived in an elementary way for Haar-random states. This is not true for random fermionic Gaussian states. In this case, we show that an analytic result can be obtained in the thermodynamic limit, based on a combination of techniques from random-matrix (RM) and large-deviation theories. We test our predictions against numerical calculations and discuss the sub-leading finite-size contributions. 

The rest of this work is organized as follows. In Sec.~\ref{sec:symmetry_resolved} we introduce the SR entanglement entropy. In Sec.~\ref{sec:haar_random} we study the ensemble of Haar-random pure states, and show that the SR Page curve can be related in an elementary way to the standard one. The ensemble of random fermionic Gaussian states is discussed in Sec.~\ref{sec:gaussian_states}. Using a combination of different techniques, we derive an explicit analytic formula valid in the thermodynamic limit, and provide exact numerical results at finite system sizes. Our conclusions are consigned to Sec.~\ref{sec:conclusions}, while some technical details of our work are reported in Appendix~\ref{app}. 

\section{Symmetry-resolved entanglement entropy}
\label{sec:symmetry_resolved}

We begin by reviewing the definition of symmetry-resolved R\'enyi and von Neumann entanglement 
entropies~\cite{laflorencie2014spin,moshe2018symmetry,xavier2018equipartition,bonsignori2019symmetry}. We consider a system $S$ in a pure state $ \rho=\ket{\psi}\bra{\psi}$ and assume there exists a $U(1)$-symmetry operator (or \emph{charge}) $\hat{ Q}$ such that, given a spatial bipartition $S=A\cup B$, we have $\hat{Q}=\hat{Q}_{A}\oplus \hat{Q}_{B}$. If $ \rho$ has a well-defined charge, i.e. $[\rho,\hat{Q}]=0$, then  $[\rho_{A},\hat{Q}_A]=0$, where $\rho_{A}={\rm tr}_{B}[\rho]$ is the reduced density matrix of the subsystem $A$. Accordingly, $\rho_{A}$ is block-diagonal, and we can write a decomposition in terms of the eigenvalues $Q$ of $\hat{Q}_A$ of the form
\be\label{eq:rhodec}
\rho_{A}=\oplus_{Q} p_{A}(Q) \rho_{A}(Q)\,,
\ee
where
\begin{eqnarray}
	 \rho_{A}(Q)=\frac{\Pi_{Q}\rho \Pi_{Q}}{p_{A}(Q)}\,,
\end{eqnarray}
with $p_{A}(Q) ={\rm tr}\left[\Pi_{Q}\rho_A\right]$, while $\Pi_Q$ is the projector onto the charge sector corresponding to eigenvalue $Q$. We use the notation $\hat{Q}$ or $\hat{Q}_A$ to distinguish the charge operators from the eigenvalues $Q$. From $\rho_{A}(Q)$, we can compute the amount of entanglement between $A$ and $B$ in each symmetry sector in terms of the SR R\'enyi entropies, defined as
\be\label{eq:SR_renyi}
S_{\alpha}(Q) \equiv \frac{1}{1-\alpha} \ln {\rm tr} [\rho_{A}^{\alpha}(Q)]\,.
\ee
The limit $\alpha \to 1$ provides the SR entanglement entropy
\be\label{eq:SRentanglement}
S_{1}(Q) \equiv-\operatorname{tr}\left[ \rho_{A}(Q) \ln \rho_{A}(Q)\right]\,.
\ee
Recalling the definition of the Neumann entanglement entropy~\cite{nielsen2002quantum}
\be
S_{1}(\rho_A)=-{\rm tr}\left[\rho_A\ln\rho_A\right]\,,
\ee
and using Eq.~\eqref{eq:rhodec}, we can decompose it in the different charge sectors as
\begin{align}\label{eq:number}
S_{1}(\rho_A)=&\sum_{Q} p_A(Q) S_{1}(Q)-\sum_{Q} p_A(Q) \ln p_A(Q)\,\nonumber\\
=:& S_{\textrm{c}}+S_{\mathrm{num}}\,,
\end{align}
where $S_{\textrm{c}}$ is known as {\it configurational entropy} and quantifies the average contribution to the total entanglement of all the charge sectors \cite{lukin2019probing,barghathi2018renyi,barghathi2019operationally}, while 
$S_{\textrm{num}}$ is called {\it number entropy} and takes into account the entanglement due to the charge fluctuations in the subsystem $A$ \cite{lukin2019probing,kiefer2020bounds,kiefer2020evidence,kiefer2021slow}.


The projection onto a given symmetry sector makes the computation of the SR entanglement entropies challenging. A simple strategy to get around this problem, which was put forward in Refs.~\cite{moshe2018symmetry,xavier2018equipartition}, is as follows. We start by computing the charged moments
\be\label{eq:charged_moments}
Z_{\alpha}(\theta) \equiv \operatorname{tr}\left[\rho_{A}^{\alpha} \mathrm{e}^{\mathrm{i} \theta \hat{Q}_{A}}\right]\,,
\ee
and their Fourier transform
\be\label{eq:fourier}
\mathcal{Z}_{\alpha}(Q)=\int_{-\pi}^{\pi} \frac{d \theta}{2 \pi} \mathrm{e}^{-\mathrm{i} Q \theta} Z_{\alpha}(\theta) \equiv \operatorname{tr}\left[\Pi_{Q} \rho_{A}^{\alpha}\right]\,.
\ee
It is easy to see that the SR R\'enyi and von Neumann entanglement entropies are then obtained as
\bea
S_{\alpha}(Q)&=\frac{1}{1-\alpha} \ln \left[\frac{\mathcal{Z}_{\alpha}(Q)}{\mathcal{Z}_{1}(Q)^{\alpha}}\right],\\ S_{1}(Q)&=-\partial_{n}\left[\frac{\mathcal{Z}_{\alpha}(Q)}{\mathcal{Z}_{1}(Q)^{\alpha}}\right]_{\alpha=1}\,,
\eea
while 
\be\label{eq:pQ_moments}
p_A(Q)=\mathcal{Z}_{1}(Q)\,.
\ee
We will make use of these formulas in numerical computations presented in Sec.~\ref{sec:gaussian_states}.

\section{Haar-random pure states}
\label{sec:haar_random}

We start by studying an ensemble of Haar-random pure states with a $U(1)$ charge. This can be realized, for instance, as the ensemble of states constructed by following up to late times a stochastic unitary dynamics with a $U(1)$ charge~\cite{rakovszky2018diffusive,khemani2018operator,piroli2020random}. We focus on 
a lattice of $L$ two-level quantum systems, associated with the Hilbert space $\mathcal{H}=\bigotimes_{j=1}^L\mathcal{H}_j$, where $\mathcal{H}_j\simeq \mathbb{C}^{2}$, and introduce the explicit symmetry operator
\begin{equation}
\hat{Q}=\frac{1}{2}\sum_{j=1}^L (\sigma^z_j+1)
\end{equation}
where $\sigma^z_j$ are Pauli matrices.  We can decompose the Hilbert space as $\mathcal{H}=\bigoplus_{M=0}^L \mathcal{H}(M)$, where $\mathcal{H}(M)\subset\mathcal{H}$ is the charge eigenspace associated with the eigenvalue $M$. Finally, we introduce the ensemble of random states $\ket{\psi}$ drawn out of the uniform Haar distribution over the set of all states in $\mathcal{H}(M)$. 

We will consider a bipartition $S=A\cup B$, where $A$ and $B$ contain $\ell$ and $L-\ell$ sites, respectively. Differently from the case studied in~\cite{page1993average}, the Hilbert space $\mathcal{H}(M)$ does not factorize into a tensor product, but we have the decomposition
\begin{equation}\label{eq:decomposition}
\mathcal{H}(M)=\bigoplus_{Q=0}^M\mathcal{H}_A(Q)\otimes \mathcal{H}_B(M-Q)\,.
\end{equation}
Here $\mathcal{H}_A(Q)$ is the eigenspace of $Q_A$ associated with eigenvalue $Q$, and similarly for $\mathcal{H}_B(M-Q)$.

Ensembles of Haar-random states over spaces admitting a decomposition of the form~\eqref{eq:decomposition} were studied recently in Ref.~\cite{bianchi2019typical}, where analytic formulae for the Page curve and its variance were derived. In fact, the results presented in Ref.~\cite{bianchi2019typical} also allow one to directly obtain the SR entanglement entropy, as we now explain. 

Given $\ket{\psi}\in\mathcal{H}(M)$, we can exploit the structure of $\mathcal{H}(M)$ in Eq.~\eqref{eq:decomposition} to write
\be\label{eq:state_decomposition}
\ket{\psi}=\sum_{Q=0}^M \sqrt{p_Q}\ket{\phi_Q} 
\ee
where $\ket{\phi_Q}\in \mathcal{H}_A(Q)\otimes\mathcal{H}_B(M-Q)$ is a normalized state, while $p_Q\geq 0$, with $\sum_Q p_Q=1$. The reduced density matrix over $A$ then reads
\begin{equation}
	\rho_A=\sum_{Q=0}^M p_Q \rho_A(Q)\,,
\end{equation}
where
\begin{equation}\label{eq:rho_q_phi}
	\rho_A(Q)={\rm tr}_B(\ket{\phi_Q}\bra{\phi_Q})\,.
\end{equation}
Here we used
\begin{equation}
	{\rm tr}_B(\ket{\phi_Q}\bra{\phi_{Q^\prime}})=\delta_{Q,Q^\prime}	\rho_A(Q)\,,
\end{equation}
which follows from the definition of $\ket{\phi_Q}$. Our goal is to compute the average entropy of the density matrix $\rho_A(Q)$ in~\eqref{eq:rho_q_phi}.

It was shown in Ref.~\cite{bianchi2019typical} that the uniform measure over $\mathcal{H}(M)$ factorizes as
\begin{equation}\label{eq:measure}
	\mathrm{d} \mu_{M}(\psi)=\mathrm{d} \nu\left(p_{0}, \ldots, p_{M}\right) \prod_{Q=0}^M \mathrm{d} \mu\left(\phi_{Q}\right)\,,
\end{equation}
where $\mathrm{d} \mu\left(\phi_{Q}\right)$ is the uniform measure over pure states in each sector $\mathcal{H}_A(Q)\otimes \mathcal{H}_B(M-Q)$, while $\mathrm{d} \nu\left(p_{0}, \ldots, p_{M}\right) $ is the multivariate beta distribution~\cite{bianchi2019typical}
\begin{multline}\label{eq:probab}
\mathrm{d} \nu\left(p_{0}, \ldots, p_{M}\right)= \\ \frac{1}{\mathcal{Z}}\delta(\sum_Qp(Q)-1)\prod_{Q}p(Q)^{d_A(Q)d_B(Q)}\mathrm{d}p(Q).
\end{multline}
The constant $\mathcal{Z}$ is introduced to normalize the measure to unity.
Eq.~\eqref{eq:measure} immediately yields the SR Page curve.  Indeed, using~\eqref{eq:rho_q_phi}, we can write the averaged bipartite R\'enyi entropies as 
\be\label{eq:renyi_inv}
S_{\alpha}(Q)=\frac{1}{1-\alpha}\int\mathrm{d} \mu\left(\phi_{Q}\right)\ln \mathrm{Tr} \rho^{\alpha}_A(Q).
\ee
Eq.~\eqref{eq:measure} implies that $\ket{\phi_Q}$ is distributed according to the invariant measure over $\mathcal{H}_A(Q)\otimes \mathcal{H}_B(M-Q)$. Therefore, we are left with the problem of computing the average entanglement entropy of a random state in a factorized Hilbert space, which is the problem studied by Page~\cite{page1993average,bianchi2021volume}. For the case of the von Neumann entanglement entropy, $\alpha=1$, we can thus apply directly Page's formula, and, for $d_{A}(Q) \leq d_{B}(Q)$, the final result reads 
\begin{align}\label{eq:SR_page_haar}
	S_{1}(Q)&=\Psi\left(d_{A}(Q) d_{B}(Q)+1\right)\nonumber\\
	&-\Psi\left(d_{B}(Q)+1\right)-\frac{d_{A}(Q)-1}{2 d_{B}(Q)}\,,
\end{align}
where $\Psi(x)=\Gamma^{\prime}(x) / \Gamma(x)$ is the digamma function (here, $\Gamma(x)$ is the Gamma function), while $d_{A}(Q)$ and $d_{B}(Q)$ are the dimensions of $\mathcal{H}_A(Q)$ and $\mathcal{H}_B(M-Q)$, namely
\begin{align}
d_A(Q)&= {\binom{\ell}{Q}}\,,\\
d_B(Q)&= {\binom{L-\ell}{M-Q}}\,.
\end{align}
For $d_{A}(Q) > d_{B}(Q)$, we can simply exploit the symmetry under exchange $A\leftrightarrow B$:  
the final result is obtained from Eq.~\eqref{eq:SR_page_haar} by exchanging $d_A(Q) \leftrightarrow d_B(Q)$.

\begin{figure}
	\includegraphics[width=0.8\linewidth]{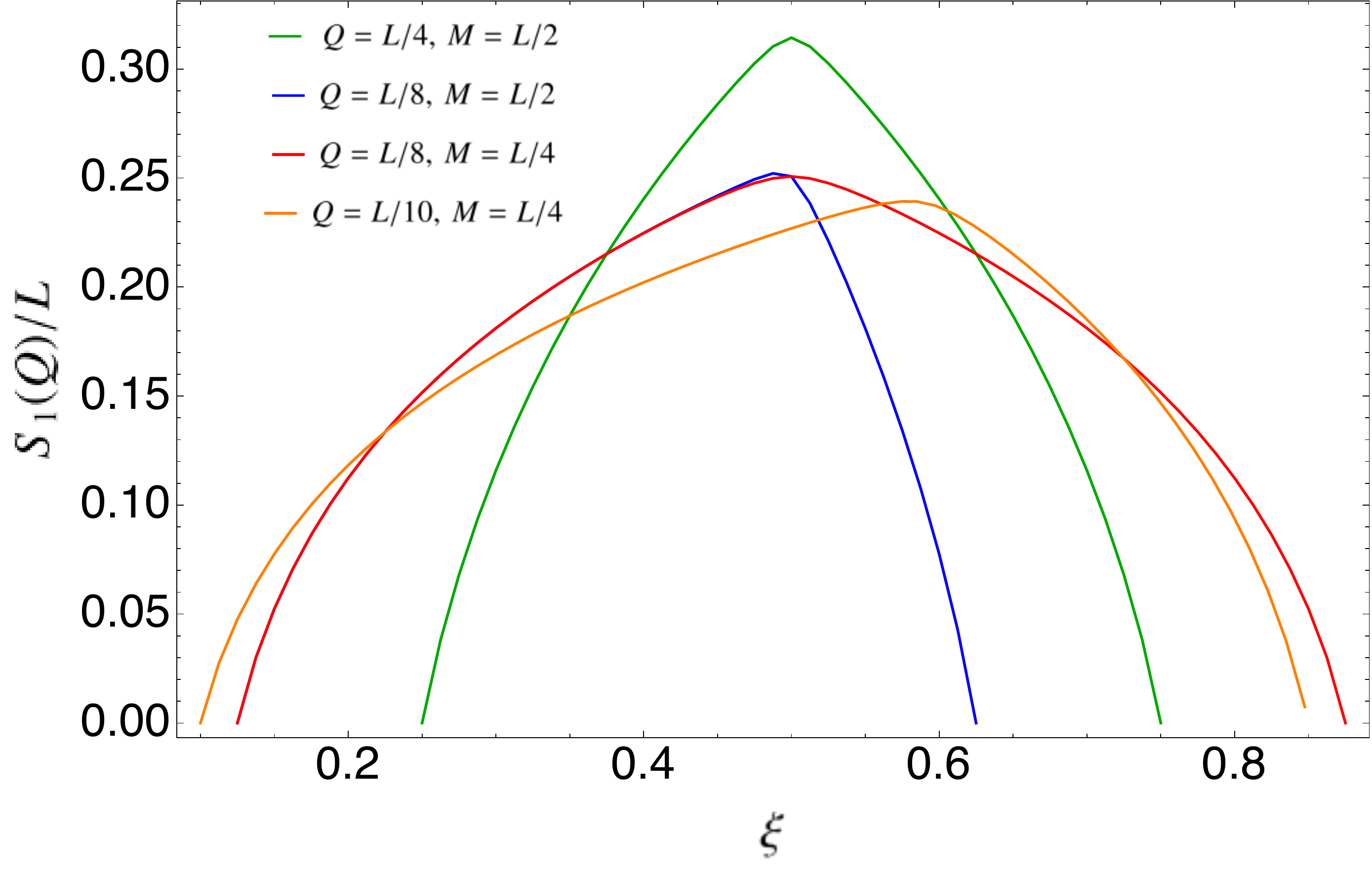}
	\caption{The average symmetry resolved entanglement entropy in Eq.~\eqref{eq:SR_page_haar} for different values of $Q,M,\xi=\ell/L$ and $L=80$. } \label{fig:nogaussian}
\end{figure}

We report the SR Page curve~\eqref{eq:SR_page_haar} in Fig.~\ref{fig:nogaussian} for different values of $L$, $M$ and $Q$. We see that the latter is in general non-symmetric with respect to the bipartition $\xi=0.5$. In addition, it vanishes for $\ell<Q$ and $\ell>L-M+Q$, as it should: in these cases, either the subspace $A$ or $B$ are not large enough to contain the available charge $Q$ or $M-Q$. Finally, in analogy to the standard case~\cite{page1993average,bianchi2021volume}, we note that the SR Page curve displays a non-analyticity when $d_{A}(Q)=d_B(Q)$.

We now consider the thermodynamic limit of Eq.~\eqref{eq:SR_page_haar}. This is performed by taking $L\to\infty$, and keeping the ratios 
\begin{align}\label{eq:rescaling}
	\xi&=\frac{\ell}{L}\,,\quad 
	m=\frac{M}{L}\,,\quad
	q=\frac{Q}{L}\,,
\end{align}
constant. In this limit, the SR entanglement entropy behaves like $S_{1}(Q)\simeq \ln d_A(Q)$, and a simple computation yields
\begin{align}
\frac{S_{1}(Q)}{L}&=\xi \ln\xi-q\ln q-(\xi-q)\ln (\xi-q)\nonumber\\
-&1/(2L)\ln\,L+o(1/L)\,.\label{eq:sq_1}
\end{align}
Importantly, we see that this formula does explicitly depend on the charge sector $Q$, breaking the equipartition at leading order $O(L)$. This is similar to what has been predicted for the SR entanglement entropy in thermodynamic states of integrable systems~\cite{piroli2022thermodynamic}. However, it is different than what has been observed so far in the study of the zero-temperature entanglement resolution, see, e.g. Ref.~\cite{xavier2018equipartition}. Our result is related to the specific order of limits we are taking, i.e. $L \to \infty$ with $q$ fixed. Conversely, we see that the next-to-leading term is independent of $Q$. Eq.~\eqref{eq:sq_1} holds for $d_{A}(Q)\leq d_{B}(Q)$, i.e.
\begin{align}\label{eq:inequality}
 \xi \ln\xi&-q\ln q\nonumber\\
 -&(\xi-q)\ln (\xi-q)
 < (1-\xi) \ln(1-\xi)\nonumber\\
 -& (m-q)\ln (m-q)\nonumber\\
 -&(1-\xi-m+q)\ln (1-\xi-m+q)\,.
\end{align}
When this inequality is not satisfied, the SR entanglement entropy is obtained from Eq.~\eqref{eq:sq_1} replacing $\xi\leftrightarrow 1-\xi$, $q\leftrightarrow m-q$. Interestingly, we see that, in the thermodynamic limit, the SR Page curve does not depend explicitly on $m$ when~\eqref{eq:inequality} holds, as it is already evident from Fig.~\ref{fig:nogaussian}. Moreover, up to subleading corrections, Eq.~\eqref{eq:sq_1} gives us the maximal value of the entropy for the bipartition $\mathcal{H}_{A}(Q)\otimes H_{B}(M-Q)$, i.e. $\ln d_A(Q)$.

\subsection{The number entropy}

The number entropy defined in Eq.~\eqref{eq:number} is obtained as an integral over the probability distribution~\eqref{eq:probab}.
The calculation is  technically rather cumbersome, but fortunately can be found in disguise in Ref.~\cite{bianchi2019typical} (see appendix B there), where it is calculated via the replica trick. Here, we only report the final result, which reads
\begin{align}\label{eq:number_fs}
S_{\mathrm{num}}&=\Psi(d_M+1)\nonumber\\
-&\sum_Q\frac{d_A(Q)d_B(Q)}{d_M}\Psi(d_A(Q)d_B(Q)+1),
\end{align}
where $d_M=\sum_Qd_A(Q)d_B(Q)$ is the dimension of the Hilbert space $\mathcal{H}(M)$.

We can also easily extract the thermodynamic limit from the finite-size result in Eq.~\eqref{eq:number_fs}.  
In order to simplify the derivation, we note that, in this limit, the averaged number entropy equals the number entropy of the averaged probability, as $\mathbb{E}[ p(Q)]$ becomes peaked around $q=m\xi$. The latter is given by $\mathbb{E}[ p(Q)]=d_A(Q)d_B(Q)/d_M$, as one can obtain by integrating  $p(Q)$ over the distribution~\eqref{eq:probab} \cite{bianchi2019typical}.
By taking the thermodynamic limit of $d_A(Q),d_B(Q)$, we find
\begin{align}
		S_{\mathrm{num}}=&-L\int_0^{\xi}dq\,\mathbb{E}[ p(Q)]\ln \mathbb{E}[ p(Q)]=0+o(L),\nonumber \\
		S_{\mathrm{c}}=&L\int_0^{\xi}dq\,\mathbb{E}[ p(Q)]S_1(Q)=\nonumber \\
		&L((m-1)\ln (1-m)-m\ln m)+o(L).
	\end{align}
The configurational entropy $S_{\mathrm{c}}$ coincides with the total entanglement entropy $S_1(\rho_A)$, satisfying the sum rule in Eq.~\eqref{eq:number}. In order to compute the first sub-leading contribution to the number entropy, we can expand $\mathbb{E}[ p(Q)]$ quadratically around $q=m\xi$, yielding
\be \label{eq:number_tl}
	S_{\mathrm{num}}=-\mathcal{N}^{-1}L \int_0^{\xi}dq \,e^{-\frac{(q-m\xi)^2}{2\sigma}}\ln (\mathcal{N}^{-1}e^{-\frac{(q-m\xi)^2}{2\sigma}}),
	\ee
	where
	\be\label{eq:variancenonG}
	\sigma=(1-m)m\xi(1-\xi)/L,\quad \mathcal{N}=L\int_0^{\xi}dq \,e^{-\frac{(q-m\xi)^2}{2\sigma}}.
	\ee
	By performing the integral in Eq.~\eqref{eq:number_tl}, we get
	\be 
	S_{\mathrm{num}}=\frac{1}{2}\ln(2\pi m(1-m)\xi(1-\xi)L)+\frac{1}{2}.
	\ee
	To summarize, we have found that the number entropy is sub-leading with respect to the total entropy and it scales as the logarithm of the variance of $p(Q)$, with a prefactor which does not depend on $m,\xi$ but it is equal to $1/2$. This can be connected to the result for conformal field theories with an internal Lie group symmetry~\cite{calabrese2021symmetry}, where the authors showed that the coefficient of such term is equal to
	half of the dimension of the group, which here is $U(1)$. Thus, despite we are not at criticality, this result is not spoiled and it seems to be even more general.

\subsection{The R\'enyi entropy}

Finally, we briefly discuss the $\alpha$-R\'enyi entropy for $\alpha>0$. For finite dimensions $d_A(Q)$, $d_B(Q)$, the presence of the logarithm in the definition~\eqref{eq:SR_renyi} makes it challenging to derive a formula analogous to~\eqref{eq:SR_page_haar} (although an exact result for the averaged moments ${\rm Tr}\rho_A^{\alpha}(Q)$ can be obtained~\cite{malacarne2002average, bianchi2019typical} and, in all random systems, they are directly related to the entanglement spectrum  \cite{Fagotti_2011}). However, in the thermodynamic limit R\'enyi entropies of Haar random states coincide with the von Neumann entropy, since such typical states, in the limit of large system size $L$,  display maximal entanglement~\cite{nadal2011statistical}. Therefore, based on our previous discussions we arrive at the simple result
	\begin{equation}
		\begin{split}
		\frac{S_{\alpha}(Q)}{L}=\,[\xi \ln\xi-q\ln q-(\xi-q)\ln (\xi-q)]+o(1).
		\end{split}
	\end{equation}

\section{Random fermionic Gaussian states}
\label{sec:gaussian_states}
We now move on to study the ensemble of fermionic random Gaussian states with a $U(1)$ symmetry~\cite{bianchi2021volume}. We take a model of $L$ fermionic modes associated with the creation and annihilation operators $c_j$, $c^\dagger_j$ with $\{c^\dagger_j,c_k\}=\delta_{j,k}$. The charge is the particle number
\be
\hat{Q}=\sum_{j=1}^Lc^\dagger_jc_j\,.
\ee
We recall that Gaussian states can be defined as the states satisfying Wick's theorem~\cite{bravyi2004lagrangian} and that, in the case where the particle number is fixed, they are completely specified by the covariance matrix
\be\label{eq:covariance_matrix}
C_{i,j}=\langle \psi|c^\dagger_i c_j |\psi\rangle\,.
\ee
The ensemble of random Gaussian states with fixed particle number $M$ can be defined by the ensemble of covariance matrices $C=V^\dagger C_0(M) V$, where $V$ is drawn out of the uniform Haar distribution over the unitary group $U(L)$ and 
\be\label{eq:c0}
C_0(M)={\rm diag}(\underbrace{1,\ldots ,1}_M, \underbrace{0,\ldots ,0}_{L-M})\,.
\ee
This distribution can be achieved as the late-time limit of stochastic Gaussian dynamics, e.g. the Quantum Symmetric Simple Exclusion process  (Q-SSEP)~\cite{bauer2017stochastic,bauer2019equilibrium,bernard2022dynamics,hruza2022dynamics}.

In this case, it is not possible to apply directly the logic of the previous section. While, given a random Gaussian state $\ket{\psi}$, we can still write the decomposition~\eqref{eq:state_decomposition}, the projected states $\ket{\phi_Q}$ are in general not Gaussian. Therefore, one cannot compute the corresponding bipartite entanglement entropy in terms of random Gaussian ensembles. Nevertheless, in the next subsection we show that the SR Page curve can be computed analytically in the thermodynamic limit~\eqref{eq:rescaling}, where we have the scaling
\be
S_{1}(Q)\sim L s_1(q)+o(L)\,.
\ee
In the next subsections we will compute $s_{1}(q)$ as a function of $\xi$, defining the SR Page curve. Our approach is based on a combination of the G\"artner-Ellis theorem~\cite{touchette2009large}, detailed in Sec.~\ref{sec:GH}, and the Coulomb-gas (CG) method of RM theory~\cite{forrester2010log,majumdar2014top}, explained in Sec.~\ref{sec:exact_SR}.

\subsection{SR entanglement from the G\"artner-Ellis theorem}
\label{sec:GH}

In principle, the SR entanglement entropy could be computed using the strategy outlined in Sec.~\ref{sec:symmetry_resolved}. However, the Fourier transform in~\eqref{eq:fourier} introduces some technical complications from the analytic point of view. In this section, we provide an alternative approach valid in the thermodynamic limit, which is based on an application of the G\"artner-Ellis theorem from large deviation theory~\cite{touchette2009large}. This method has been recently introduced in Ref.~\cite{piroli2022thermodynamic} to compute the SR entanglement entropy of thermodynamic states in quantum integrable models.

Let us denote by $\rho$ the density matrix of the subsystem $A$. First, using
\be
[\rho,\Pi_Q]=0\,,
\ee
and the fact that $\Pi_Q^\alpha=\Pi_Q$ for any power $\alpha$, we can write
\be\label{eq:p-s_decomposition}
{\rm tr} [\rho^{\alpha}(Q)]=p_\alpha(Q) \frac{{\rm tr}[\rho^\alpha]}{(p_1(Q))^\alpha}\,,
\ee
where we have introduced
\be
p_\alpha(Q)=\frac{{\rm tr} [\Pi_Q\rho^{\alpha}]}{{\rm tr}[\rho^\alpha]}\,.
\ee
Therefore, the averaged SR R\'enyi entropy is
\begin{align}\label{eq:exchange_log}
\mathbb{E}[S_\alpha(Q)]&=\frac{1}{1-\alpha}\left(\mathbb{E}\left[ \ln p_{\alpha}(Q)\right]-\alpha \mathbb{E}\left[ \ln  p_{1}(Q)\right]\right)\nonumber\\
&+\mathbb{E}\left[ S_{\alpha}\right]\,,
\end{align}
where $S_{\alpha}$ is the standard (non-resolved) R\'enyi entropy, while $\mathbb{E}[\cdot]$ now denotes the average over the ensemble of random Gaussian states.

In order to proceed, we make the assumption that
\be\label{eq:self_averaging_1}
\mathbb{E}\left[ \ln p_{\alpha}(Q)\right]=  \ln  \mathbb{E}\left[ p_{\alpha}(Q)\right]+o(L)\,,
\ee
namely that, up to sub-leading-order corrections, we can bring the average ``inside the logarithm''. We have tested numerically the validity of this assumption, cf.~\ref{sec:numerical_results}, which can be justified invoking the concentration of measure for fermionic random Gaussian states~\cite{bianchi2021page,bernard2021entanglement}. Now, it is immediate to see that $\mathbb{E}\left[p_\alpha(Q)\right]\geq 0$ and $\sum_{Q} \mathbb{E}\left[p_\alpha(Q)\right]=1$. Therefore, $\mathbb{E}\left[p_\alpha(Q)\right]$ can be interpreted  as a probability distribution. Setting $q=Q/L$, we expect on physical grounds that $\mathbb{E}\left[p_\alpha(Q)\right]$ follows a large deviation principle in the large-$L$ limit, that is,
\be
\mathbb{E}\left[p_\alpha(Q)\right]\sim e^{-I_{\alpha}(q)L}\,,
\ee
where $I_{\alpha}(q)$ is referred to as the \emph{rate function}~\cite{touchette2009large}. To compute it, we define the generating function
\be\label{eq:generating_function}
G_\alpha(w)=\mathbb{E}\left[\frac{{\rm tr}\left[ \rho^{\alpha} e^{w \hat{Q}}\right]}{{\rm tr}[\rho^\alpha]}\right]\,,\qquad w\in \mathbb{R}\,,
\ee
and
\be\label{eq:f_function}
f_\alpha(w)=\lim_{L\to\infty}\frac{1}{L}\ln G_{\alpha}(w)\,.
\ee
The G\"artner–Ellis theorem~\cite{touchette2009large} states that we can compute $I_{\alpha}(q)$ as the Legendre transform of  $f_\alpha(w)$, namely
\be\label{eq:rate_function_alpha}
I_\alpha(q)=w_{\alpha, q} q-f_\alpha\left(w_{\alpha, q}\right)\,,
\ee
where $w_{\alpha, q}$ is determined by the condition
\be
\left.\frac{d}{d w}(f_\alpha(w)-w q)\right|_{w=w_{\alpha, q}}=0\,.
\label{eq:sqdefinition_alpha}
\ee
Finally, introducing the density of SR R\'enyi entropy
\be
s_\alpha(q)=\lim_{L\to\infty} \frac{S_\alpha(q L)}{L}\,,
\ee
and using~\eqref{eq:exchange_log}, we obtain
\be\label{eq:final_result_alpha}
s_\alpha(q)=s_\alpha+\frac{1}{1-\alpha}\left[-I_{\alpha}(q)+\alpha I_{1}(q)\right]\,,
\ee
where 
\be 
s_{\alpha} =  \lim_{L\to\infty}\frac{\mathbb{E}\left[ S_{\alpha}\right]}{L}\,,
\ee
is the density of the standard (non-resolved) Rényi entropy. The von Neumann SR entanglement entropy is obtained taking the limit $\alpha\to1$ 
\be\label{eq:final_result_vN}
s_1(q)=s_1+I_1(q)+\frac{{\rm d} I_\alpha(q)}{{\rm d}\alpha}\Big|_{\alpha=1}\,.
\ee

\subsection{The Coulomb-gas approach}
\label{sec:exact_SR}

In order to obtain the SR entanglement entropy, we need to compute the function $f_\alpha(w)$ in Eq.~\eqref{eq:f_function}. To this end, we make use of the CG approach~\cite{forrester2010log,majumdar2014top}. In the context of fermionic random Gaussian states, this method has been recently applied in Refs.~\cite{liu2018quantum,pengfei2020subsystem,bernard2021entanglement} to compute average bipartite entanglement entropies and their large deviations. Here, we briefly review the aspects of these works directly relevant for our purposes.

As mentioned, the covariance matrix~\eqref{eq:covariance_matrix} contains full information about the corresponding Gaussian state, encoding also its bipartite entanglement entropy~\cite{vidal2003entanglement}. In particular, given a region $A$ containing $\ell$ sites, and denoting by $C^{A}$ the $\ell\times\ell$ matrix with $C^{A}_{i,j}=C_{i,j}$ for $i,j\in A$, the corresponding R\'enyi entropy reads
\be\label{eq:entropy_eigen}
S_{\alpha}=\frac1{1-\alpha}\sum_{j=1}^\ell\ln\left[\lambda_j^\alpha+(1-\lambda_j)^\alpha\right]\,.
\ee
Here $\{\lambda_j\}_{j=1}^\ell$ are the eigenvalues of $C^{A}$,  satisfying $0\leq \lambda_j\leq 1$. When $C$ is sampled according to the invariant measure discussed above, the eigenvalues $\lambda_j$ are random variables. Their  probability distribution, $P[\{\lambda_j\}]$, is known~\cite{forrester2010log}, and takes the form
\be\label{eq:jacobi_spectrum}
P[\left\{\lambda_{i}\right\}]=\frac{1}{\mathcal{N}}\prod_{j<k }\left|\lambda_{j}-\lambda_{k}\right|^{2} \prod_{i=1}^{\ell} \lambda_{i}^{M-\ell}(1-\lambda_i)^{L-\ell-M}\,,
\ee
where $\mathcal{N}$ is a normalization constant. This distribution defines the $\beta$-Jacobi ensemble (with $\beta=2$), see Refs.~\cite{vivo2008distributions,vivo2010probability,damle2011phase} for applications in different physical contexts~.

In principle, Eq.~\eqref{eq:jacobi_spectrum} allows one to compute the expectation value of arbitrary functions of the eigenvalues. However, for finite $\ell$ this is often complicated, as averages involve integrals in $\ell$-dimensional spaces. When $\ell\to\infty$, the problem can be simplified using a standard method of RM theory, consisting in a mapping between the eigenvalues $\lambda_j$ and a Coulomb gas of
repulsive point charges~\cite{forrester2010log}. In order to see how it works, we consider a function $g(\{\lambda_k\})$ and write its expectation value as 
\be\label{eq:functional}
\mathbb{E}[g(\{\lambda_k\})]=\frac{1}{\mathcal{N}}\int_0^1 \prod_{j=1}^\ell {\rm d}\lambda_j\, e^{- \ell^2E[\{\lambda_j\}]}g(\{\lambda_k\}\,,
\ee 
with 
\begin{align*}
	E[\{\lambda_{i}\}]&=-\frac{2}{\ell^2}\sum_{i<j} \ln |\lambda_{i}-\lambda_{j}|
	-\frac{(M-\ell)}{\ell^2} \sum_{i} \ln \lambda_{i}
	\nonumber\\
	-&\frac{(L -M-\ell)}{\ell^2} \sum_{i} \ln \left(1-\lambda_i\right)\,.
\end{align*}
Within the CG formalism, $E[\{\lambda_{i}\}]$ is interpreted as the energy of a gas of charged particles with coordinates $\lambda_j\in[0,1]$ and subject to an external potential. The integral~\eqref{eq:functional} is the thermal partition function for the CG. In the large-$\ell$ limit, the configuration of the particles may be described in terms of the normalized density $\rho(\lambda)=\ell^{-1}\sum_j\delta(\lambda-\lambda_i)$, and the multiple integral in Eq.~\eqref{eq:functional} can be cast into a functional integral over all possible densities $\rho(\lambda)$, i.e. 
\be\label{eq:functional_integral}
\mathbb{E}[g]= \int \mathcal{D}\rho\, e^{-\ell^2 E[\rho]}g[\rho]\,.
\ee
To the leading order in $\ell$~\footnote{When we replace the multiple integral with the functional integral, we need to take into account the Jacobian $J[\rho]$ of the change of coordinates. It can be argued that $J[\rho]\sim e^{O(\ell)}$,  so we can neglect it at the leading order in $\ell$~\cite{extreme2008dean}.}, $E[\rho]$  reads
\begin{align}
	E[\rho]=&-\int_0^1 {\rm d}\lambda\int_0^1{\rm d}\mu\, \rho(\lambda)\rho(\mu)\ln |\lambda-\mu|\nonumber\\
	+&\int_0^1 {\rm d}\lambda\rho(\lambda)V(\lambda)+u\left\{\int_0^1{\rm d}\lambda\,\rho(\lambda)-1\right\}\,,
\end{align}
where we introduced the Lagrange multiplier $u$ enforcing normalization, and the effective potential
\begin{align}\label{eq:effective_potential}
	V(\lambda)=&-\left(\frac{m}{\xi}-1 \right)\ln \lambda-\left(\frac{1-m}{\xi}-1 \right)\ln\left (1-\lambda\right)\,,
\end{align}
where $m$, $\xi$ are the density of fermions and the rescaled interval length introduced in Eq.~\eqref{eq:rescaling}.

The functional integral~\eqref{eq:functional_integral} can be computed via the saddle-point method, and the average is dominated by the typical distribution function $\rho^\ast(\lambda)$ satisfying $\delta E[\rho]/\delta \rho|_{\rho=\rho^\ast}=0$. The solution is known~\cite{forrester2012large,ramli2012spectral} and reads
\be\label{eq:average_distribution}
\rho^\ast(\lambda)=\frac{1}{2 \pi \xi} \frac{\sqrt{\left(\nu_{+}-\lambda\right)\left(\lambda-\nu_{-}\right)}}{\lambda(1-\lambda)}\,,
\ee
with $\lambda\in[\nu_{-},\nu_+]$ and $\nu_{\pm}=[\sqrt{m(1-\xi)} \pm \sqrt{\xi(1-m)}]^{2}$. Eq.~\eqref{eq:average_distribution} allows us to compute directly, in the thermodynamic limit, averages of extensive quantities which can be written as sums over the eigenvalues, by replacing them with integrals over $\lambda\in[\nu_{-},\nu_+]$. For example, the averaged R\'enyi entropy~\eqref{eq:entropy_eigen} can be computed as 
\be\label{eq:integral}
\mathbb{E}[S_\alpha]=\ell\int_{\nu_-}^{\nu_+} {\rm d}\lambda\rho^\ast(\lambda) \sigma_\alpha(\lambda)\,,
\ee
where $\sigma_{\alpha} (\lambda)=\frac{1}{1-\alpha} \ln \left[\lambda^{\alpha}+(1-\lambda)^{\alpha}\right]$, see Ref.~\cite{pengfei2020subsystem} for an explicit expression of this integral.

\subsection{Exact SR Page curves}
\label{sec:final_result}

We finally combine the techniques outlined in the previous subsections, and obtain an analytic result for the SR entanglement entropy. Our starting point is the computation of $f_{\alpha}(w)$ defined in Eq.~\eqref{eq:f_function}. Once again, we can use a typicality argument and exchange the order of the expectation value and the logarithm in Eqs.~\eqref{eq:exchange_log}, ~\eqref{eq:f_function}. We are left with the task of computing
\begin{equation}
f_{\alpha}(w)=\mathbb{E}[\ln{\rm tr}( \rho^{\alpha} e^{w \hat{Q}})]-\mathbb{E}[\ln{\rm tr}(\rho^{\alpha})]\,.
\end{equation}
Following the previous subsection, we write the rhs in terms of the eigenvalues $\lambda_j$ of the reduced covariance matrix $C^{A}$. In particular, we have
\be\label{eq:cm_eigen}
\ln{\rm tr}( \rho^{\alpha} e^{w \hat{Q}})=\sum_{j=1}^\ell\ln\left[e^{w}\lambda_j^\alpha+(1-\lambda_j)^\alpha\right]\,.
\ee

In the thermodynamic limit $\ell \to \infty$, we can apply the  
CG approach and obtain
\begin{multline}
\lim_{L\to\infty}\frac{\mathbb{E}[\ln{\rm tr}( \rho^{\alpha} e^{w \hat{Q}})]}{L}=\\\xi \int_{\nu_-}^{\nu_+} d\lambda\rho^*(\lambda)\ln (e^{w}\lambda^{\alpha}+(1-\lambda)^{\alpha}).
\end{multline}
Consequently, $I_{\alpha}(q)$ in Eq.~\eqref{eq:rate_function_alpha} reads
\begin{multline}\label{eq:GEn}
I_{\alpha}(q)=w_{\alpha,q} q - \\ \xi \int_{\nu_-}^{\nu_+}d\lambda\rho^*(\lambda)\ln (e^{w_{\alpha,q}}\lambda^{\alpha}+(1-\lambda)^{\alpha})+\\ \xi \int_{\nu_-}^{\nu_+} d\lambda\rho^*(\lambda)\ln (\lambda^{\alpha}+(1-\lambda)^{\alpha})].
\end{multline}
The value of $w_{\alpha,q}$ is fixed by Eq.~\eqref{eq:sqdefinition_alpha}, which can be rewritten as
\begin{equation}\label{eq:tointerpolate}
q=\xi\int_{\nu_-}^{\nu_+} d\lambda\rho^*(\lambda)\frac{e^{w_{\alpha,q}}\lambda^{\alpha}}{e^{w_{\alpha,q}}\lambda^{\alpha}+(1-\lambda)^{\alpha}}\,.
\end{equation}

\begin{figure*}
	{\includegraphics[width=0.47\linewidth]{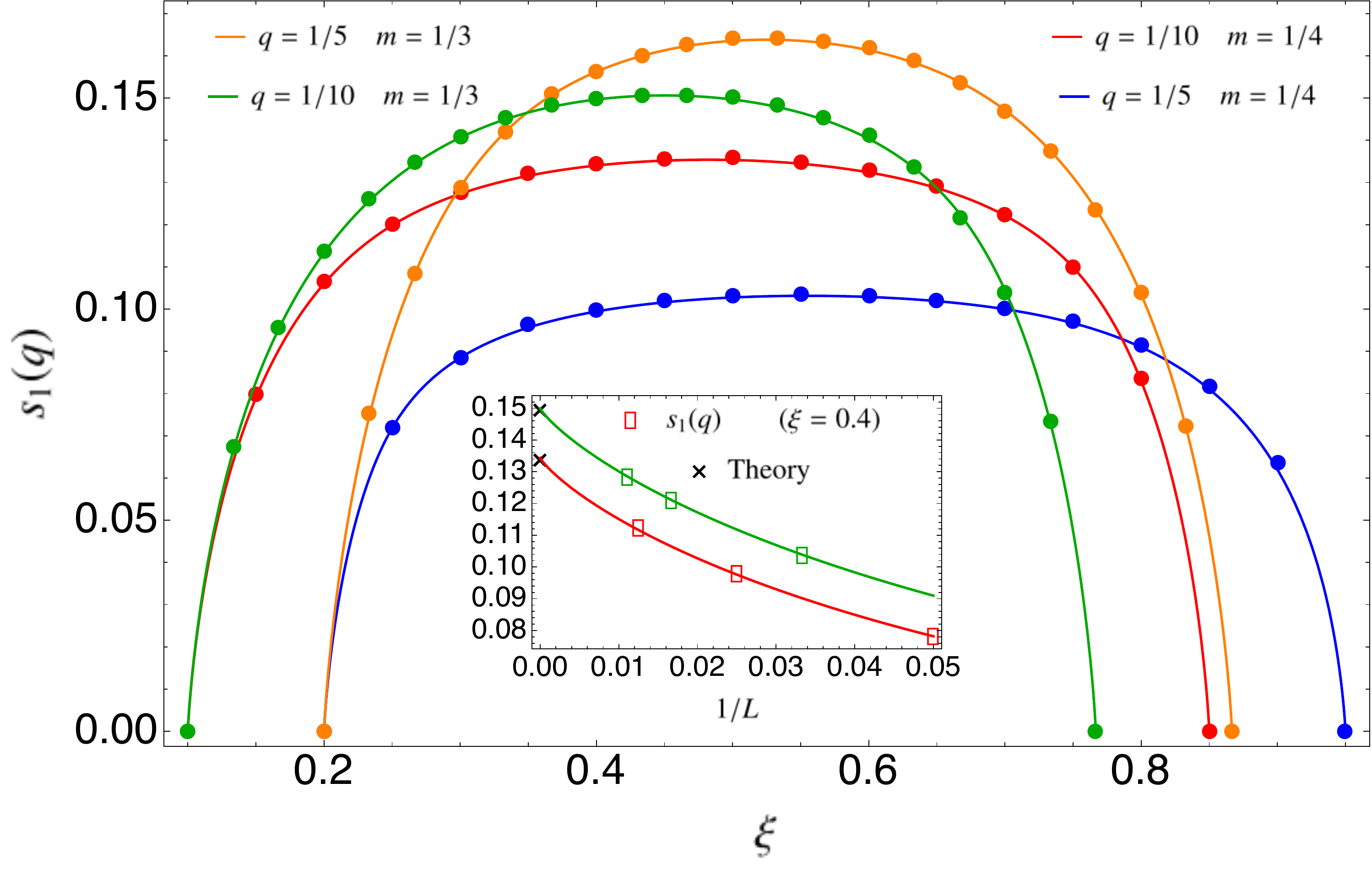}}
	\subfigure
	{\includegraphics[width=0.47\linewidth]{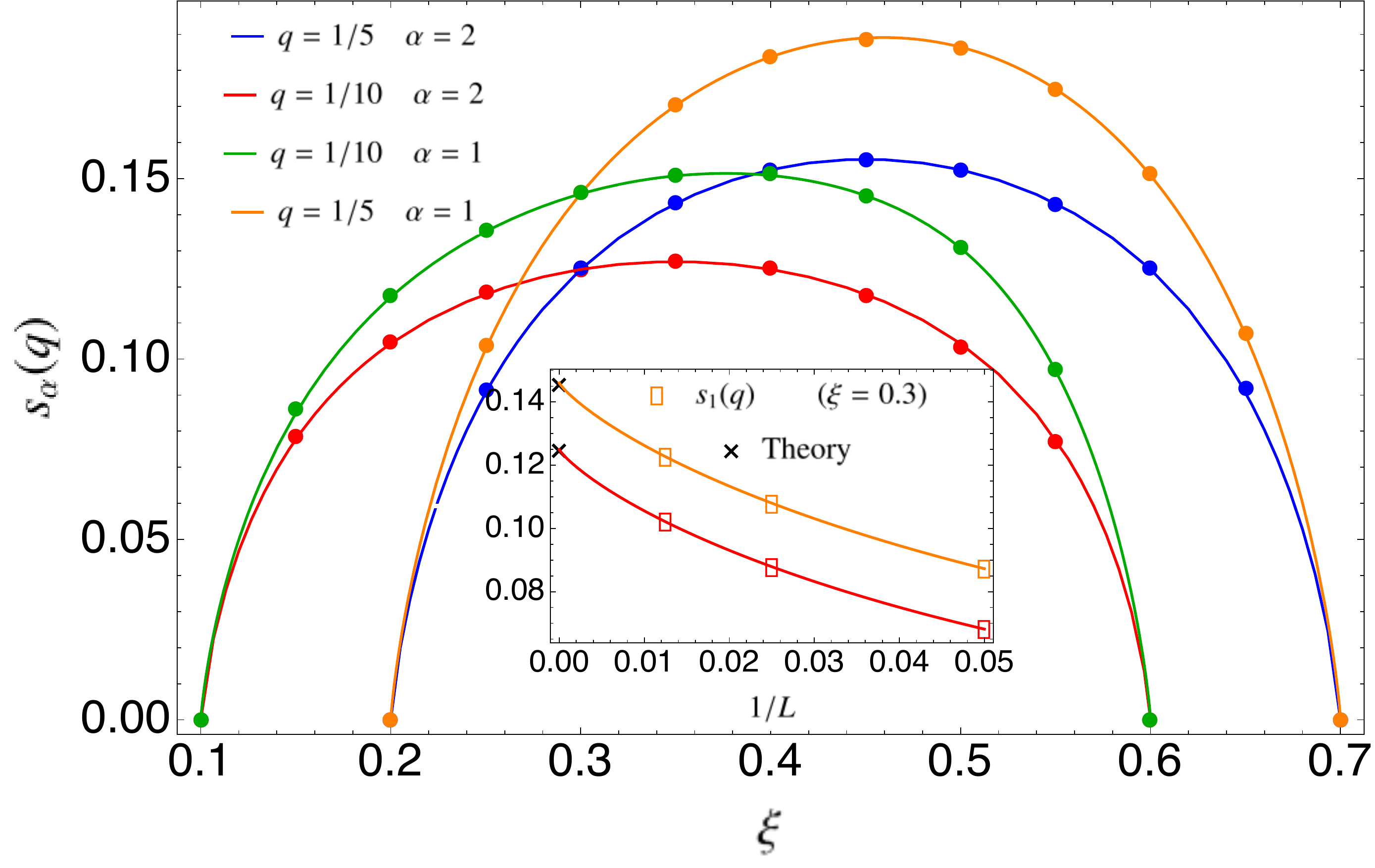}}
	\caption{Comparison between the asymptotic results derived in subsection~\ref{sec:final_result} and the exact values of the average entanglement entropy $s_{\alpha}(q)$ computed numerically for $L\to \infty$ (symbols). They have been obtained by using the extrapolation form $s_1(q)=a-1/(2L)\ln L+b/L$. Insets show data for different values of $L$. The solid line for $\alpha=1$ corresponds to Eq.~\eqref{eq:generic_m} and its extension to $\xi>m$ according to Eq.~\eqref{eq:geqm}--\eqref{eq:extension_3}.} \label{fig:gaussian}
\end{figure*}

Plugging $w_{\alpha,q}$ into the expression for $I_{\alpha}(q)$ in Eq.~\eqref{eq:GEn}, we get the density of SR R\'enyi entropy $s_{\alpha}(q)$ in Eq.~\eqref{eq:final_result_alpha}. From Eq.~\eqref{eq:tointerpolate}, we see that $w_{\alpha,q}$ can be obtained by inverting the equation numerically, after evaluating simple integrals. We followed this procedure to generate plots of the function $s_{\alpha}(q)$ for $\alpha>1$, as reported in Fig.~\ref{fig:gaussian}. For generic $\alpha$,  we were not able to find an analytical expression of $w_{\alpha,q}$. However, Eq.~\eqref{eq:tointerpolate} can be computed explicitly and inverted in the limit $\alpha \to 1$ (see Appendix~\ref{app} for the details).  In this case, we find
\begin{equation}\label{eq:w1q}
\begin{split}
w_{1,q}=&\ln\left[ \frac{q(-1+m-q+\xi)}{(m-q)(q-\xi)} \right],
\end{split}
\end{equation}
and, according to Eq.~\eqref{eq:final_result_vN}, we are left with
\begin{widetext}
\begin{align}\label{eq:spn11}
s_1(q)&=-(w_{1,q}-w'_{1,q}) q+\xi\int_{\nu_-}^{\nu_+} d\lambda\rho^*(\lambda)\ln (e^{w_{1,q}}\lambda+(1-\lambda))\nonumber\\
&-\xi\int_{\nu_-}^{\nu_+} d\lambda\rho^*(\lambda) \frac{(1-\lambda)\ln(1-\lambda)+e^{w_{1,q}}\lambda(\ln \lambda)}{(e^{w_{1,q}}-1)\lambda+1}-\int_{\nu_-}^{\nu_+} \frac{d\lambda}{2\pi}\frac{\sqrt{(\lambda-\nu_-)(\nu_+-\lambda)}}{(1-\lambda)}\frac{e^{w_{1,q}}w'_{1,q}}{(e^{w_{1,q}}-1)\lambda+1}.
\end{align} 
In order to simplify our computation, we observe that, from Eq.~\eqref{eq:tointerpolate},
\begin{equation}
	\int_{\nu_-}^{\nu_+} \frac{d\lambda}{2\pi}\frac{\sqrt{(\lambda-\nu_-)(\nu_+-\lambda)}}{(1-\lambda)}\frac{e^{w_{1,q}}}{(e^{w_{1,q}}-1)\lambda+1}=q,
\end{equation}
which can be straightforwardly substituted in the last term of Eq.~\eqref{eq:spn11} giving $-w'_{1,q}q$. This implies that we do not need the compute $w'_{1,q}$, as it cancels out in Eq.~\eqref{eq:spn11}. The remaining integrals can be solved following the techniques used in Ref.~\cite{pengfei2020subsystem}, cf. Appendix~\ref{app} for further details about their solutions. Putting all together, we arrive at the final result (for $q\leq \xi\leq m$) 
\begin{align}\label{eq:generic_m}
s_1(q)&=\frac{(-1 + m) q (-1 + \xi) \ln(1 - m)}{q - m \xi} + 
	\frac{m (q - \xi) (1 - \xi) \ln m}{q - m \xi} +\frac{ 
		m (q - \xi) (-1 + \xi) \ln(m-q)}{q -m\xi} \nonumber \\
	&- 
	q \ln q -\frac{ (1 - m) q (1 - \xi) \ln(1 - m + q - \xi)}{q -
		m \xi}+ \xi \ln\xi+ (q - \xi) \ln( \xi-q )\,.
\end{align}
\end{widetext}
We remark that the equipartition of the entanglement entropy is explicitly broken also in this case.
For $\xi>m$, we can obtain the SR entanglement entropy using the symmetries of the problem~\cite{bianchi2021volume}. In particular, it is not difficult to show that the following relations hold (making explicit the dependence on $\xi$ and $m$)
\begin{align}
\label{eq:geqm}
s_1(m;q; \xi)&=s_1(\xi;q; m),\quad   m< \xi \leq 0.5 \\
s_1(m;q; \xi)&= s_1(1-\xi;m-q; m), \ 0.5< \xi \leq 1-m 
\end{align}
and
\begin{equation}\label{eq:extension_3}
s_1(m;q; \xi)= s_1(m;m-q; 1-\xi)\,,
\end{equation}
for $  1-m< \xi \leq 1-m+q$.
Eq.~\eqref{eq:generic_m} is our main result, which will be further discussed in the next subsection.

Finally, let us compute the averaged number entropy. To this end, we assume again that, in the thermodynamic limit, it equals the number entropy of the averaged probability. Using  $\mathbb{E}[ p(Q)]= e^{-L I_{1}(q)}\mathcal{N}$ [where $\mathcal{N}$ is a normalization constant], we have that $\mathbb{E}[ p(Q)]$ is peaked around $q=m\xi$, and to the leading order in $L$ we have 
\begin{align}
S_{\mathrm{num}}=&-L\int_0^{\xi}dq\,\mathbb{E}[ p(Q)]\ln \mathbb{E}[ p(Q)]=0+o(L),\nonumber \\
S_{\mathrm{c}}=&L^2\,\int_0^{\xi}dq\,\mathbb{E}[ p(Q)] s_1(q)=L \left\{ \, (\xi-1)\ln (1-\xi)\right.\nonumber \\
+&\left.\xi[(m-1)\ln(1-m)-m\ln m -1]\right\}+o(L).
\end{align}
This implies that the sum rule in Eq. \eqref{eq:number} is satisfied and the number entropy is sub-leading with respect to the total entropy. In order to find the first sub-leading term, we can expand $\mathbb{E}[ p(Q)]$ quadratically around $q=m\xi$, and we obtain
\be 
S_{\mathrm{num}}=-\mathcal{N}^{-1}\, L\int_0^{\xi}dq \,e^{-\frac{(q-m\xi)^2}{2\sigma}}\ln (\mathcal{N}^{-1}\,e^{-\frac{(q-m\xi)^2}{2\sigma}}),
\ee
where
\be
\sigma=(1-m)m\xi(1-\xi)/L,\quad \mathcal{N}=L\int_0^{\xi}dq \,e^{-\frac{(q-m\xi)^2}{2\sigma}}.
\ee
By comparing this result with Eq.~\eqref{eq:variancenonG}, this computation shows that the expressions for $\mathbb{E}[ p(Q)]$ for Haar-random and random Gaussian states are the same in the thermodynamic limit. Therefore, remarkably, we get the same result for the first sub-leading correction to the number entropy 
\be \label{eq:number_gaussian}
S_{\mathrm{num}}=\frac{1}{2}\ln(2\pi m(1-m)\xi(1-\xi)L)+\frac{1}{2}.
\ee

\subsection{Numerical results}
\label{sec:numerical_results}

We have tested Eq.~\eqref{eq:generic_m} against numerical computations. We have sampled the ensemble of random Gaussian states at finite system sizes by generating covariance matrices $V^\dagger C_0(M)V$, where $C_0(M)$ is defined in Eq.~\eqref{eq:c0} and $V$ is drawn from the uniform distribution over $U(L)$. For each Gaussian state, we have computed the SR entanglement entropies following the method outlined in Sec.~\ref{sec:symmetry_resolved}, using that the charged moments can be expressed in terms of the eigenvalues of the correlation matrix $C^A$, i.e.
\begin{equation}
Z_{\alpha}(\theta)=\prod_{j=1}^{\ell}[\lambda_j^{\alpha}e^{i\theta}+(1-\lambda_j)^{\alpha}]\,.
\end{equation}
For fixed $m$, $\xi$, we have considered $10^3$ random samples. The SR entanglement entropies are obtained by taking the mean value over them. We have repeated this procedure for different system sizes, $L$, and we have extrapolated the data at finite $L$ in order to recover the thermodynamic limit $L \to \infty$. This allows us to compare the numerical data against the analytical predictions found in the previous subsection. 

In Fig.~\ref{fig:gaussian}, we show the comparison between the extrapolated data in the thermodynamic limit and the density of the SR entropies for different values of $q, m, \xi, \alpha$. In all cases, the error associated with the finite-number of samples and the fitting procedure is not visible in the scales of the plot, and is therefore omitted. As in the case of Haar-random pure states, we see that the SR Page curves are not symmetric with respect to $\xi=0.5$, and are vanishing for $\xi<q$ and $\xi>1-m+q$. The numerical results are found to be in excellent agreement with our analytic predictions.

It is interesting to discuss the finite-size corrections and our fitting procedure. Interestingly, our numerical results convincingly show that the sub-leading corrections to $s_{\alpha}(q)$ are proportional to $(1/2)\ln L/L$, independent of $q$, $m$ and $\xi$, cf. the inset of Fig.~\ref{fig:gaussian}. Accordingly, we have performed a fit of our data against the function
\begin{equation}
	s_1(q)=a-1/(2L)\ln L+b/L\,.
\end{equation}
We note that the sub-leading behavior is exactly the same as that for Haar random pure states, cf. Eq.~\eqref{eq:sq_1}. We note also that the sub-leading behavior $-(1/2)\ln L$ to the SR entanglement entropies $S_{\alpha}(q)$ have been observed in other contexts when the total entropies are extensive, see for instance~Ref. \cite{parez2021quasiparticle,scopa2022exact}

\begin{figure}
	\includegraphics[width=0.8\linewidth]{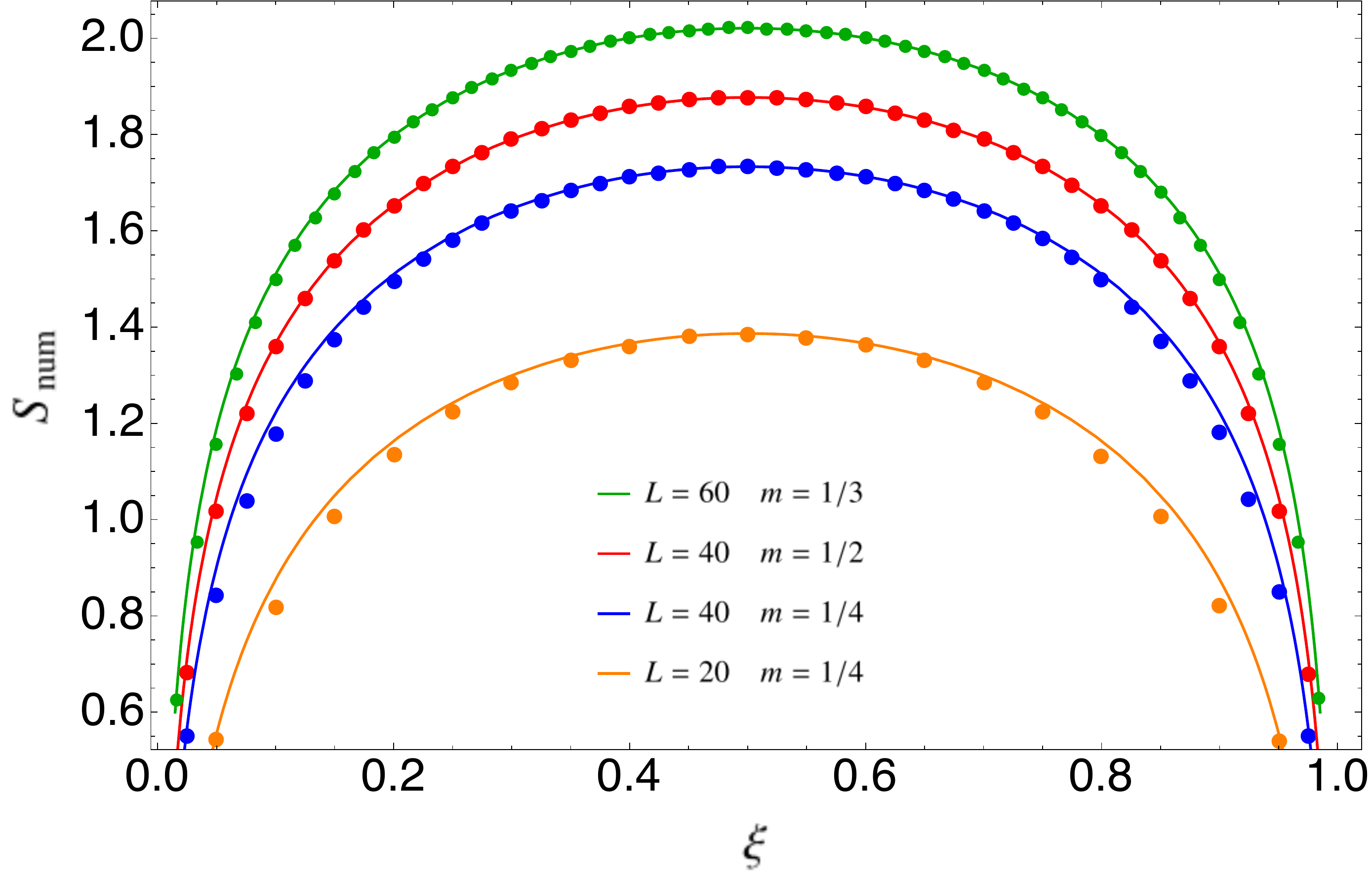}
	\caption{The average number entropy in Eq.~\eqref{eq:number_gaussian} for different values of $m,\xi,L$. } \label{fig:number}
\end{figure}

Finally, we tested the first sub-leading correction to the number entropy described in~Eq. \eqref{eq:number_gaussian} in Fig.~\ref{fig:number}. We find that the agreement improves as the system size $L$ increases, being excellent for $L=60$.

\section{Conclusions}
\label{sec:conclusions}

In this work, we have considered the computation of the SR Page curves for two important statistical ensembles with a $U(1)$-symmetry: Haar-random pure states and random fermionic Gaussian states. In the former case, we have shown how an exact result can be obtained in an elementary way for finite systems and in the thermodynamic limit. This is not true for fermionic Gaussian states. Still, we were able to compute the corresponding SR Page curve in the thermodynamic limit. Our main technical tools have been the Coulomb gas method and  the G\"artner-Ellis theorem. We expect that our approach could allow for the computation of SR R\'enyi entropies in other situations where the latter are extensive.

One could wonder whether similar calculations can be extended to ensembles of random bosonic Gaussian states. As pointed out in Ref.~\cite{bianchi2021volume}, however, the only bosonic Gaussian state with a fixed particle number is the vacuum state. In addition, while one could choose an ensemble of bosonic Gaussian states with a fixed average number of particles, the very definition of SR entanglement entropy requires an exact conservation law. Obviously, one could also consider complex bosons and in that case the $U(1)$ symmetry is easily implemented and the SR entropy is always well defined, as in the field theory case~\cite{murciano2020entanglement}. We are not aware of any work in this direction.

Our results are expected to be relevant in a number of cases. As already pointed out, fermionic random Gaussian states are realized as the large-time limit of sufficiently generic stochastic Gaussian dynamics, as proven rigorously, for instance, in the Q-SSEP~\cite{bauer2017stochastic,bauer2019equilibrium,bernard2022dynamics}. Our formulae immediately apply in those cases. In addition, they could be used as a comparison in more general situations, for instance to evaluate non-Gaussian effects in the SR entanglement entropy of thermodynamic states in interacting integrable systems~\cite{piroli2022thermodynamic}. 

Finally our results might be relevant for the information paradox in charged black holes in analogy to the role played by the original the Page curve for the neutral case. In particular, the results for replica wormholes \cite{almheiri2020,stanford2022} should have a counterpart also for charged black holes.


\section*{Acknowledgments} 

PC and SM acknowledge support from ERC under Consolidator grant number 771536 (NEMO).

\bigskip
\appendix
\section{Details on analytic derivations}\label{app}
In this Appendix, we sketch some details required to obtain the results reported in Sec.~\ref{sec:final_result}. 
The first integral we want to solve is
\begin{equation}\label{eq:toinvert}
q=\int_{\nu_-}^{\nu_+} d\lambda\frac{1}{2 \pi } \frac{\sqrt{\left(\nu_{+}-\lambda\right)\left(\lambda-\nu_{-}\right)}}{(1-\lambda)}\frac{e^{w_{\alpha,q}}}{(e^{w_{\alpha,q}}-1)\lambda+1}\,.
\end{equation}
Performing the change of variable $\lambda'=1-\lambda$, one can rewrite the previous integral as
\begin{equation}
(1-e^{-w_{\alpha,q}})^{-1}\int_{1-\nu_+}^{1-\nu_-} d\lambda \frac{\sqrt{\left(\nu_{+}-1+\lambda\right)\left(-\lambda+1-\nu_{-}\right)}}{2 \pi\lambda(\lambda-(1-e^{-w_{\alpha,q}})^{-1})}\,.
\end{equation}
This can be solved by means of the identity~\cite{pengfei2020subsystem}
\begin{align}
&c\displaystyle\int_a^b \frac{dx}{2\pi}\frac{\sqrt{(b-x)(x-a)}}{x(c-x)}\nonumber\\
&=\frac{c}{2}\left(1-\frac{\sqrt{ab}+\sqrt{(c-a)(c-b)}}{c} \right),
\end{align}
after the substitution 
\begin{align}
a &=1 -\nu_+\,,\\
b &=1-\nu_-\,,\\
c &=(1-e^{-w_{\alpha,q}})^{-1}\,.
\end{align}
Finally, we can invert the resulting analytic formula in Eq.~\eqref{eq:toinvert} and find $e^{w_{1,q}}$ as a function of $m,q, \xi$, yielding Eq.~\eqref{eq:w1q}. 

Next, we explain the result in Eq.~\eqref{eq:generic_m}. We rewrite Eq.~\eqref{eq:spn11} as
\begin{equation}
s_1(q)=-w_{1,q}q\\+J_1-J_2-J_3,
\end{equation}
where 
\begin{align}
J_1=&\xi\int_{\nu_-}^{\nu_+} d\lambda\rho^*(\lambda)\ln (e^{w_{1,q}}\lambda+(1-\lambda)),\\
J_2=&\xi\int_{\nu_-}^{\nu_+} d\lambda\rho^*(\lambda) \frac{(1-\lambda)\ln(1-\lambda)}{(e^{w_{1,q}}-1)\lambda+1},\\
J_3=&\xi\int_{\nu_-}^{\nu_+} d\lambda\rho^*(\lambda) \frac{e^{w_{1,q}}\lambda(\ln \lambda)}{(e^{w_{1,q}}-1)\lambda+1}.
\end{align}
For the first integral, we can perform the change of variables $\lambda'=(1-e^{w_{1,q}})\lambda $, yielding
\begin{multline}\label{eq:J1}
J_1=\int_{(1-e^{w_{1,q}})\nu_-}^{(1-e^{w_{1,q}})\nu_+} \frac{d\lambda}{2\pi}\\ \frac{\sqrt{\left(\nu_{+}(1-e^{w_{1,q}})-\lambda\right)\left(\lambda-\nu_{-}(1-e^{w_{1,q}})\right)}}{\lambda(1-e^{w_{1,q}}-\lambda)} \ln (1-\lambda).
\end{multline}
By means of simple manipulations, the second one can cast in the form
\begin{align}\label{eq:J2}
J_2&=(1-e^{w_{1,q}})^{-1}\nonumber\\
&\int_{\nu_-}^{\nu_+} \frac{d\lambda}{2\pi}
\frac{\sqrt{\left(\nu_{+}-\lambda\right)\left(\lambda-\nu_{-}\right)}}{\lambda(1-e^{w_{1,q}}-\lambda)} \ln (1-\lambda)\,. 
\end{align}
Finally, for the third one, the change of variables $\lambda'=1-\lambda$ yields 
\begin{multline}\label{eq:J3}
J_3= (1-e^{-w_{1,q}})\int_{1-\nu_+}^{1-\nu_-} \frac{d\lambda}{2\pi}\\\frac{\sqrt{(\lambda-1+\nu_-)(1-\nu_+-\lambda)}}{\lambda(1-e^{-w_{1,q}}-\lambda)} \ln(1-\lambda).
\end{multline}
The integrals featuring in the definitions of $J_1$, $J_2$ and $J_3$ can be solved using the following identity~\cite{pengfei2020subsystem}
\begin{widetext}
\begin{align}\label{eq:ref}
&\displaystyle\int_a^b \frac{dx}{2\pi}\frac{\sqrt{(b-x)(x-a)}}{x(c-x)}\ln(1-x)=\nonumber\\
&\frac{1}{2}\left(1-\frac{\sqrt{ab}}{c}-\frac{\sqrt{(c-a)(c-b)}}{c} \right)\ln(1-a)+I(1)
-\frac{\sqrt{ab}}{c}I\left(\sqrt{\frac{a}{b}}\right)-\frac{\sqrt{(c-a)(c-b)}}{c}I\left(\sqrt{\frac{c-a}{c-b}} \right),
\end{align}
where $0<a<b<|c|$ or $c<b<a<0$, and
\begin{equation}
I(\eta)=\ln \frac{1+\sqrt{\frac{b-1}{a-1}}\eta}{1+\eta}.
\end{equation}
The integrals in Eqs.~\eqref{eq:J1}, \eqref{eq:J2}, \eqref{eq:J3} can be identified (up to a global prefactor) with the one in Eq.~\eqref{eq:ref}, provided that the following substitutions for $a$, $b$, $c$ in Eq.~\eqref{eq:ref} are made, respectively:
\begin{align}
 a & =\nu_-(1 - e^{w_{1,q}})\,,\qquad  b =\nu_+(1 - e^{w_{1,q}})\,, \qquad c=1 - e^{w_{1,q}}\,,  \\
 a & =\nu_-\,,\qquad b =\nu_+\,,\qquad c =(1 - e^{w_{1,q}})^{-1}\,,\\
 a & =1 -\nu_+\,,\qquad b =1-\nu_-\,, \qquad c =(1 - e^{-w_{1,q}})^{-1}\,.
\end{align}
\end{widetext}
Let us notice that the expression of $w_{1,q}$ for $\xi <m$ is greater than 0, so $e^{w_{1,q}}>1$ and the inequalities $0<a<b<|c|$ or $c<b<a<0$ are satisfied.
Using the expression for $\nu_{\pm}$ and $w_{1,q}$ reported in Eq.~\eqref{eq:w1q} and summing the contribution of each integral, we find the result in Eq.~\eqref{eq:generic_m}.

\bibliography{bibliography}

\end{document}